\documentstyle[12pt]{article}
\begin{document}
\thispagestyle{empty}
\def\thebibliography#1{\section*{}\list
{\arabic{enumi}.}{\settowidth\labelwidth{#1.}\leftmargin\labelwidth
\advance\leftmargin\labelsep
\usecounter{enumi}}
\def\newblock{\hskip .11em plus .33em minus -.07em}
\sloppy
\sfcode`\.=1000\relax}
\let\endthebibliography=\endlist

\begin{titlepage}
\begin{center}
\hspace*{7cm} Preprint IFUNAM\\
\hspace*{7cm}  FT-93-024\\
\hspace*{7cm} August 1993 \\
\vspace*{2cm}
{\Large{\bf Relativistic Covariant Equal-Time  Equation\\ \vskip1mm
For Quark-Diquark System}}\\
\vspace*{4mm}
{\tt VALERI V. DVOEGLAZOV}$^{*, \,\dagger}$\\
\vskip1mm
{\it Departamento de F\'{\i}sica Te\'orica\\
Instituto de F\'{\i}sica, UNAM, Apartado Postal 20-364\\
 01000 Mexico, D. F.\,\,  MEXICO}\\
\vskip2mm {\tt and}\\
{\tt SERGEI V. KHUDYAKOV}$^{*, \,\ddagger}$\\
\vskip1mm
{\it  Laboratory of Theoretical Physics, JINR\\ Head Post Office, P. O. Box
79\\
Moscow 101000 (Dubna) \,\, RUSSIA}\\
\vskip2mm {\tt and}\\
{\tt SVYATOSLAV B. SALGANIK}\\
\vskip1mm
{\it Department of Theoretical} \& {\it Nuclear Physics\\
Saratov State University, Astrakhanskaya str. , 83\\
Saratov  410071  RUSSIA} \\
\end{center}
\vspace*{1cm}
\small{
{\tt ABSTRACT.}  Relativistic three-dimensional quasipotential (equal-time)
equations are considered, which describe  bound states of  fermion and
boson of spin $S=0$ or $S=1$. The spin structure of the interaction
quasipotentials in such systems is studied, and corresponding partial-wave
equation for the simplest case is obtained. Such  equations can be used
in calculations of energy spectra, decay rates and structure functions
of quark-diquark systems (nucleons and their resonances) as well as
for description of ($\pi\mu$)- atom.}

\vspace*{1cm}
\vspace*{1cm}


\vspace*{-5mm}
\noindent
-----------------------------------------------------------------------------\\
\noindent
\footnotesize{
$^{*}$ On leave from: {\it Dept.Theor.} \& {\it Nucl. Phys.,
   Saratov State University and Sci.} \& {\it Tech. Center for  Control
and Use of Physical Fields and Radiations, Astrakhanskaya str. , 83,\,\,
   Saratov 410071 RUSSIA}\\
$^{\dagger}$ Email: valeri@ifunam.ifisicacu.unam.mx,
dvoeglazov@main1.jinr.dubna.su\\
$^{\ddagger}$ Email: khud@theor.jinrc.dubna.su,
vapr@scnit.saratov.su}
\end{titlepage}

\vspace*{7mm}
\section{\bf  Introduction}

\hspace*{8mm}The concept of constituent diquarks has been introduced in 1966,
Ref.~\cite{Diquark}. In a three-quark system spin-spin interaction
can lead to an existence of the
short-range correlations in two-quark subsystems~\cite{Narodets}, which are
comparable in strength to the $\bar q q$- attraction inside mesons. There are
experimental evidences for the diquark correlations in
baryons~\cite{Buck}\footnote{Let us still mention that another point of view is
presented  in some papers~\cite{Lein}.}. Scalar diquarks are mentioned
in~\cite{Fred} to be energetically favored.  Moreover,  in a series of recent
papers~\cite{Neubert} it was shown that
the concept of diquarks as effective degrees of freedom arising as a result
of such correlations has  important meaning for descriptions of nonleptonic
weak decays at low energies. Scalar diquarks also arise
in superstring-inspired models~\cite{Conto}. Therefore,  nucleon can be
interpreted as the quark-diquark bound state and  described on the base of
well-known methods for a solution of two-body
problem. In connection with that we can mention the papers of Lichtenberg
with collaborators~\cite{Licht} and Efimov's group~\cite{Efimov}\footnote{See
also the recent reviews~\cite{Szc,Richard}.}. There are also many speculations
concerning
"diquonia" (diquark-antidiquark bound system) and "dibaryons",  which
are rather based on the radical point of view of considering diquarks as
elementary constituents.

Secondly, we would like to note that now the new trend in  physics  of
elementary  particles
connected with the researches of  properties  of  so-called  exotic
atoms has been developed. Such systems   represent  the  atoms  in
which one of  electrons is replaced by  an   elementary particle \cite{Kiril}.
The first works devoted to the consideration of these systems appeared in the
forties~\cite{Wheeler}.  In
the  middle of the  70--s  the bound state of $\pi$ -- meson and muon
\cite{Sch},
which also can be interpreted as the exotic atom, has been   experimentally
observed.
It   deserves   to  emphasize  that the  main  properties of ($\pi\mu$)- atom
have been studied
theoretically  in  Ref.~\cite{Nemenov} even before its experimental discovery.
In these papers the attention  has been paid  at the possibilities  of
exploring features of $\pi$--meson  by means of
experimental investigation of the composite system of meson and lepton  ( one
could find the following works in Ref. ~\cite{Bar}).

In the works ~\cite{Karim} the influence of the
relativistic  effects in the description of ($\pi\mu$)-atom was under
consideration.  For this purpose the equal-time  quasipotential approach
suggested
by  Logunov  and  Tavkhelidze in  Ref. ~\cite{Logunov} has been used for the
description of these  bound
states  on the base of  quantum  field  methods.

In the presented paper we employ the quasipotential approach, the Kadyshevsky's
version~\cite{Kadysh},  to the model in which the nucleon is considered to be a
bound state of a quark of spin $1/2$ and a diquark, whose spin is $S=0$ or
$S=1$.
We   are  interested in   the
spin    structure   of  the  quasipotentials   for interaction  between
fermion (e.g., quark or $\mu$--meson)  and
(pseudo) scalar particle  (e.g., diquark or $\pi$--meson) as well as between
fermion and vector particle which is described by Joos-Weinberg's
formalism~\cite{Weinberg}.  We also find out the form of  the quasipotential in
the partial-wave equation  for  ($\pi\mu$)- atom
\footnote{The analogous problem for two spinor particle system has been solved
in~\cite{Dvoeglaz3}.} and
propose the ways of numerical solutions of the above-mentioned
equations.

\section {\bf Equation For the Wave Function of the Composite System Formed by
Fermion and $S=0$ Boson}

\hspace*{8mm}The   quasipotential  equation  for the  wave   function of the
composite
system consisting of fermion and spinless boson has been obtained in
\cite{Link}
\begin{eqnarray}
2 \Delta^{0}_{p, m_2 \lambda_P} (M - \Delta^{0}_{p,m_1
\lambda_P}-\Delta^{0}_{p,m_2 \lambda_P})
\Phi_{\sigma}(\vec \Delta_{p,\lambda_P}) = \nonumber\\
= \frac{1}{(2\pi)^3} \sum_{\nu} \int \frac{d^{3} \vec\Delta_{k,\lambda_P}}
{2\Delta^{0}_{k,m_1 \lambda_P}} V^{\nu}_{\sigma}
(\vec \Delta_{p,\lambda_P}; \vec \Delta_{k,\lambda_P}) \Phi_{\nu}
(\vec \Delta_{k,\lambda_P}).
\end{eqnarray}

The quasipotential  $\hat V$  coincides  with  the  scattering  amplitude
of    muon    on    pion  in the  first  approximation  in   the
coupling  constant.   Covariantly    defined     in     the
c.m.s.   4-momentum   of    particles         is
presented  by  the following  formulas ~\cite{Link}-\cite{Skach}
\footnote{We omit the circles above $\vec p$ and $\vec k$ in the following,
implying still the covariant generalizations of the usual momenta.}
($\lambda_P =\frac{p_1 + p_2}{\sqrt{(p_1 + p_2)^2}}$):
\begin{equation}
\vec\Delta_{p, \lambda_P}=\vec \Delta_{p, m_1 \lambda_P} = (L^{-1}_{\lambda_P}
\vec p_1) = \vec{p}_{1}-\frac{\vec P}{M}
(p_{1}^{0} - \frac{\vec P\vec p_1}{P_{0} + M})=-\vec \Delta_{p, m_2 \lambda_P}
\equiv \breve{\vec p};
\end{equation}

\begin{equation}
\Delta^{0}_{p, m_j \lambda_P}= \sqrt{\vec\Delta^2_{p, \lambda_P} + m^2_j}\equiv
\breve{p_{1}^{0}};  j=1, 2
\end{equation}
Here, $M$ is the mass of bound system, $m_1$ is muon  mass, $m_2$ is pion mass,
$L^{-1}_{\lambda_P}$ is the
matrix  of  the  Lorentz  boost  from  the  system with $4$--momentum $P_{\mu}$
and $4$--velocity $\lambda^{\mu}_P \equiv P^{\mu} / \sqrt{P^2}$
to the rest system, $L^{-1}_{\lambda_P} P = (\vec M ,\vec 0)$. The covariant 4-
momentum of the particle after interaction ($\Delta^0_{k, m_j \lambda_P}$ and
$\vec \Delta_{k, \lambda_P}$) is defined similarly.

In Ref. ~\cite{Link2}  expression  for  the   quasipotential is chosen in the
form:
\begin{eqnarray}\label{eq:4}
\lefteqn{\hat V^{(2)\,\nu}_{\quad\sigma}(\vec\Delta_{p,\lambda}
;\vec\Delta_{k,\lambda}) =\sum_{pol. inds.} D^{+\quad(S=1/2)}_{\sigma\sigma_p}
\left \{V^{-1}(\Lambda_P, p_1)\right \}\times}\nonumber\\
&\times&V_0(\vec\Delta_{k,\lambda}-\vec\Delta_{p,\lambda})
\overline{u}(\vec\Delta_{p,\lambda};
\sigma_p)\gamma_{\mu}u(\vec\Delta_{k,\lambda};\nu_p)(\tilde\Delta_{p,\lambda}
+\tilde\Delta_{k,\lambda})^{\mu}\times\nonumber\\
&\times& D^{(S=1/2)}_{\nu_p \nu_k} \left \{V^{-1} (\Lambda_{p_1}, k_1)\right \}
D^{(S=1/2)}_{\nu_k \nu}\left \{V^{-1} (\Lambda_P, k_1)\right \},
\end{eqnarray}
where $V_0$ is the  local  part   of the quasipotential  corresponding  to the
one-boson  exchange,  $\Delta^{\mu}_{p,\lambda_P}=(\Delta^0_{p,m_1\lambda_P};
\vec\Delta_{p,\lambda_P})$; $\tilde\Delta^{\mu}_{p,
\lambda_P}=(\Delta^0_{p,m_2\lambda_P};
-\vec\Delta_{p, \lambda_P})$, $D^{(S=1/2)}$ are Wigner functions.

Let us  rewrite  (\ref{eq:4})  in   more   details    using    the
results  of  Ref. ~\cite{Skach}.  We employ the expression of Ref.~\cite{Skach}
for the 4- current
\begin{equation}\label{eq:current}
j^{\mu}_{\sigma_p\nu_p} (\vec p, \vec k)=\overline {u}(\vec p,
\sigma_p)\gamma^{\mu} u(\vec k, \nu_p)=
\frac{2}{\sqrt{2m(\Delta^0 +m)}}
\xi^{*}_{\sigma_p} \left [p^{\mu}(\Delta_0 +m)
+ 2W^{\mu}(\vec p)(\vec\sigma \vec\Delta)\right ]\xi_{\nu_p},
\end{equation}
with
\begin{equation}
\vec\Delta\equiv\vec k (-) \vec p=(L^{-1}_{\vec p} \vec k);\quad
\Delta^0\equiv
\sqrt{\vec\Delta^2+m^2}=\frac{p^{\mu} k_{\mu}}{m}
\end{equation}
is  momentum transfer in the  Lobachevsky space.
$W^{\mu}(\vec p)$ is  the vector  of relativistic
spin (Pauli--Lyubansky--Shirokov vector, $p_{\mu}W^{\mu}(\vec p)=0$ ;
$k_{\mu}W^{\mu}(\vec p)= -\frac{m}{2}(\vec\sigma\vec\Delta)$).
We "reseted" the polarization indices to a single momentum, e.g., $\vec p$ as
earlier~\cite{ Dvoeglaz3},\cite{Skach}-\cite{Dvoeglaz}.
As a result we  obtain\footnote{Let us mention that the another version of the
quasipotential approach, based on two-time Green function formalism, has been
used in~\cite{Kapsh}. This way leads to  the quasipotentials depending on the
total energy of bound system.}
\begin{eqnarray}\label{eq:qq}
\hat{V}^{(2)}(\vec k, \vec p)&=&\frac{2}{\sqrt{2 m_1 (\Delta^0_1 + m_1)}}
\left \{[p_{1}^{0} (p_{1}^{0}+ p_{2}^{0}+k_{1}^{0}+ k_{2}^{0})-2 m^2_1]
(\Delta^0_1+m_1)+\right.\nonumber\\
&+&\left.(\vec p \vec\Delta_1)(p_{1}^{0}+ p_{2}^{0}+ k_{1}^{0}+ k_{2}^{0}) +
i\vec\sigma[\vec p \vec\Delta_1]
(p_{1}^{0} + p_{2}^{0}+ k_{1}^{0}+  k_{2}^{0})\right \} V_0(\vec k (-) \vec p).
\end{eqnarray}
After   transition  to  the nonrelativistic limit\footnote{More exactly, to the
quasirelativistic limit similarly to the use of $1/c^2$ expansion  in the Breit
equation for two spinor particle interaction, see ~\cite{Landau}.}, one can
see  that  the  quasipotential  (\ref{eq:qq})  transforms  to  the   following
form  ($\vec \Delta_{{\cal E}} =\vec k -\vec p$):
\begin{eqnarray}\label{eq:nonrel}
V^{(2)}_{nonrel.}(\vec k,\vec p)&=&-g^2_V\frac{4m_1m_2}{\vec\Delta^2_{{\cal
E}}}+g^2_V \frac{1}{c^2}\left (1+\frac{m_2}{2m_1}\right )
-g^2_V \frac{1}{c^2}\left (2+\frac{m_1}{m_2}
+\frac{m_2}{m_1}\right )\frac{\vec p\,^2+\vec k\,^2}{\vec\Delta^2_{{\cal
E}}}-\nonumber\\
&-&g_V^2 \frac{1}{c^2}\frac{m_2}{m_1}\frac{(\vec k\,^2-\vec p\,^2)^2}{\vec
\Delta_{\cal E}^4}-g_V^2 \frac{1}{c^2}\left (1+\frac{m_2}{m_1}\right
)\frac{2i\vec\sigma[\vec p\vec\Delta^2_{{\cal E}}]}
{\vec\Delta^2_{\cal E}},
\end{eqnarray}
provided that the local part of the quasipotential is chosen in the form
\begin{equation}
V_0(\vec k(-)\vec
p)=\frac{g^2_V}{(p_1-k_1)^2}=-\frac{g^2_V}{2m_1(\Delta_1^0-m_1)},
\end{equation}
$g_V$ is  coupling constant for the quark - vector boson and diquark -
vector boson interactions.
After some calculations we obtain the matrix elements of
the quasipotential  (\ref{eq:qq}), \, $\hat V^{\nu}_{\sigma}(\vec k,\vec p)$.
They  can  be  written  in the  following  form:
\begin{eqnarray}
V^{-1/2}_{1/2}&=&\frac{2V_0(\vec k(-)\vec p)}{\sqrt{2m_1(\Delta_1^0 +m_1)}}
(p_{1}^{0}+p_{2}^{0}+k_{1}^{0}+k_{2}^{0})(i\eta_1+\eta_2),\\
V^{1/2}_{-1/2}&=&\frac{2V_0(\vec k(-)\vec p)}{\sqrt{2m_1(\Delta_1^0+m_1)}}
(p_{1}^{0}+p_{2}^{0}+k_{1}^{0}+k_{2}^{0})(i\eta_1-\eta_2),\\
V^{\,1/2}_{\,1/2}&=&\frac{2V_0(\vec k(-)\vec p)}{\sqrt {2m_1(\Delta_1^0+m_1)}}
\Bigl[\bigl(p_{1}^{0}(p_{1}^{0}+p_{2}^{0}+k_{1}^{0}+k_{2}^{0})-2m^2_1\bigr)
(\Delta_1^0+m_1)+\nonumber\\
&+&(p_{1}^{0}+p_{2}^{0}+k_{1}^{0}+k_{2}^{0})
(\vec p\vec\Delta_1)+i\eta_3(p_{1}^{0}+p_{2}^{0}+k_{1}^{0}+k_{2}^{0})\Bigr],\\
V^{-1/2}_{-1/2}&=&\frac{2V_0(\vec k(-)\vec p)}{\sqrt {2m_1(\Delta_1^0+m_1)}}
\Bigl[\bigl(p_{1}^{0}(p_{1}^{0}+p_{2}^{0}+k_{1}^{0}+k_{2}^{0})-2m^2_1\bigr)
(\Delta_1^0+m_1)+\nonumber\\
&+&(p_{1}^{0}+p_{2}^{0}+k_{1}^{0}+k_{2}^{0})
(\vec p\vec\Delta_1)-i\eta_3(p_{1}^{0}+p_{2}^{0}+k_{1}^{0}+k_{2}^{0})\Bigr],
\end{eqnarray}
where
$\vec\eta=[\vec p\vec\Delta]$.

Expanding  the  wave   function   and  the  quasipotential    in
partial  waves,  we   obtain the  system  of  partial
equations~\cite{Dvoeglaz}
\begin{equation}
2p_{2}^{0}(M - p_{1}^{0} -p_{2}^{0}) \frac{1}{p} \Psi_{Jl}(p)=
\frac{1}{2\pi}\int^{\infty}_{0}\frac{k dk}{k_{1}^{0}}
\sum_{l^{\prime}} V^{J}_{ll^{\prime}} (k,p)\Psi_{Jl^{\prime}}(k),
\end{equation}
where  $J=\vert l-1/2 \vert , l+1/2$, $k=\vert\vec\Delta_{k, \lambda}\vert$;
$p=\vert\vec\Delta_{p, \lambda}\vert$.
The coefficients $V^{J}_{ll^{\prime}} (k,p)$ can be found by the formula:
\begin{eqnarray}
V^{J}_{ll^{\prime}}(k,p)&=&\sum_{M, \sigma, \nu} \int^{\pi}_{0}
sin\theta_{p}
d\theta_{p} \int^{2\pi}_{0} d\phi_{p} \int^{\pi}_{0} sin\theta_{k} d\theta_{k}
\int^{2\pi}_{0} d\phi_{k}\times\nonumber\\
&\times&\left [\Omega^{*(1/2)}_{JlM}(\vec n_{p})\right ]^{\sigma}
V^{\nu}_{\sigma}(\vec k (-) \vec p; \vec p)\left
[\Omega^{1/2}_{Jl^{\prime}M}(\vec n_{k})\right ]_{\nu}.
\end{eqnarray}
Here,  $\vec n_{p} = \frac{\vec p}{\vert \vec p \vert}$,
$\vec n_{k} = \frac{\vec k}{\vert\vec k\vert}$, $\theta_{p}, \phi_{p}$ are
angular  coordinates  of  the vector $\vec n_{p}$;
$\theta_{k}, \phi_{k}$ are  angular  coordinates  of  the vector $\vec n_{k}$;
$\Omega^{(1/2)}_{JlM} (\vec n)$ are  spherical  spinors.

Let us choose the coordinate system in such a way that the
vector $\vec n_{p}$ is aligned to the $OZ$ axis and the vector $\vec n_{k}$
lies
in the $XZ$ plane.
The  results  of  calculations   can   be   expressed in the  integrals
$I^{(l)}_1$ and $I^{(l)}_2$
\begin{eqnarray}
\lefteqn{V^{l\pm
\frac{1}{2}}_{ll}(p,k)=-(J+\frac{1}{2})\frac{g_V^2}{m_1\sqrt{2pk}}
\left \{\Bigl [
p_{1}^{0}(p_{1}^{0}+p_{2}^{0}+k_{1}^{0}+k_{2}^{0})
\gamma^{+}-\right.}\nonumber\\
&-&\left.2m^2_1(\gamma^{+}-\gamma^{-})-
\frac{p}{k}k_{1}^{0}(p_{1}^{0}+
p_{2}^{0}+k_{1}^{0}+k_{2}^{0})\Bigr ]
I^{(l)}_1-2m^2_1I^{(l)}_2\right \}\mp\nonumber\\
&\mp&(J+\frac{1}{2})\frac{g_V^2}{\sqrt{2pk}}
(p_{1}^{0}+p_{2}^{0}+k_{1}^{0}+k_{2}^{0})
\frac{l(l+1)}{2l+1}(I^{(l-1)}_1-I^{(l+1)}_1)
(1-\delta_{l0}).
\end{eqnarray}
The matrix element for the $\Delta l=\pm 1$  transition drops out
\begin{equation}
V^{l\pm \frac{1}{2}}_{l,l\pm 1}(p,k)=0.
\end{equation}
Here, $\gamma^{+}=\frac{p_{1}^{0} k_{1}^{0}+m^2_1}{pk}$,
$\gamma^{-}=\frac{p_{1}^{0}k_{1}^{0}-m^2_1}{pk}$ and
\begin{eqnarray}\label{eq:int}
I^{(l)}_1&=&\int^1_{-1}dz\frac{P_{l}(z)}{(\gamma^{-}-z)\sqrt{\gamma^{+}-z}},\\
I^{(l)}_2&=&\int^1_{-1}dz\frac{P_l(z)}{\sqrt{\gamma^{+}-z}},
\end{eqnarray}
$P_l(z)$ is  Legendre polynomial of the first kind.

The value of the second integral can be taken from Ref.~\cite[p. 822]{Grad}
\begin{equation}
I^{(l)}_2=\frac{2^{1-l}}{2l+1}(\sqrt{\gamma^{+}+1}-\sqrt{\gamma^{+}-1})^{2l+1}.
\end{equation}

In cases of  the low angular momenta  ($l=0, 1, 2$),   the first integral can
be directly
calculated from (\ref{eq:int})
\begin{eqnarray}
I^{(0)}_1&=&\frac{1}{\sqrt{\gamma^{+}-\gamma^{-}}} \ln\left[
\frac{(\sqrt{\gamma^{+}+1}-
\sqrt{\gamma^{+}-\gamma^{-}})
(\sqrt{\gamma^{+}-1}+
\sqrt{\gamma^{+}-\gamma^{-}})}{(\sqrt{\gamma^{+}+1}
+\sqrt{\gamma^{+}-\gamma^{-}})
(\sqrt{\gamma^{+}-1}-\sqrt{\gamma^{+}-\gamma^{-}})}
\right],\\
I^{(1)}_1&=&\gamma^{-} I^{(0)}_1-2(\sqrt{\gamma^{+}+1}-\sqrt{\gamma^{+}-1}),\\
I^{(2)}_1&=&({3\over 2}\gamma^{-^{\, 2}}-{1\over 2})
I^{(0)}_1-(\gamma^{+}-\sqrt{\gamma^{+^{\,2}}-1}+3\gamma^{-})
(\sqrt{\gamma^{+}+1}-\sqrt{\gamma^{+}-1}).
\end{eqnarray}

However, calculation  of the first integral for an arbitrary $l$, the orbital
quantum number, is highly complicated and the result seems
not to be expressed  in the known special functions. See {\tt Appendix} for
some speculations in connection with this subject.

\section{\bf Equation For the Wave Function of the Composite System Formed by
Fermion and $S=1$ Boson}

\hspace*{8mm}The equation for the equal-time WF of the composite system of
fermion and $S=1$ boson is analogous to the one presented in the Section II.
\begin{eqnarray}
2 \Delta^{0}_{p,m_2 \lambda_P} (M - \Delta^{0}_{p,m_1
\lambda_P}-\Delta^{0}_{p,m_2 \lambda_P})
\Phi_{\sigma_1 \sigma_2}(\vec \Delta_{p,\lambda_P}) = \nonumber\\
= \frac{1}{(2\pi)^3} \sum_{\nu_1 \nu_2} \int \frac{d^{3}
\vec\Delta_{k,\lambda_P}}
{2\Delta^{0}_{k,m_1 \lambda_P}} V^{\nu_1 \nu_2}_{\sigma_1 \sigma_2}
(\vec \Delta_{p,\lambda_P}; \vec \Delta_{k,\lambda_P}) \Phi_{\nu_1 \nu_2}
(\vec \Delta_{k,\lambda_P})
\end{eqnarray}
It is
obvious that in this case the quasipotential in the momentum representation
does have the additional terms (which are responsible for the spin-spin
interaction, the tensor interaction and the squared spin-orbit interaction)
comparing to the case of fermion - $S=0$ boson~\cite{Dvoeglaz2}.

Following to the technique of "resetting" the polarization indices,
we get analogously to the Section II\,\,\footnote{ Let us mention that
$\xi_{\sigma_{1p}}$,
$\xi_{\nu_{1p}}$ are  usual Pauli two-component spinors normalized by
equation $\xi^{*}_{\sigma}\xi^{\nu}=\delta_{\sigma}^{\nu}$
and $\xi_{\sigma_{2p}}$,
$\xi_{\nu_{2p}}$ are  3- component analogues of Pauli spinors for $S=1$
particle.}
\begin{eqnarray}
\lefteqn{<p_1, p_2; \sigma_1, \sigma_2\vert \hat V^{(2)}\vert k_1,  k_2;
\nu_1, \nu_2> =}\nonumber\\
&=&\sum_{pol. inds.}
D^{+\quad (S=1/2)}_{\sigma_1\sigma_{1p}} \left \{V^{- 1} (\Lambda_ P,
p_1)\right \}
 D^{+\quad (S=1)}_{\sigma_2\sigma_{2p}}
\left \{V^{-1}(\Lambda_ P, p_2)\right \}\times\nonumber\\
&\times&V^{\nu_{1p}\nu_{2p}}_{\sigma
_{1p}\sigma_{ 2p}}(\vec k(-)\vec p, \vec p) D^{(S=1/2)}_{\nu_{1p}\nu_{1k}}\left
\{
V^{-1} (\Lambda_{p_1}, k_1)\right \} D^{(S=1/2)}_{\nu_{1k}\nu_1}\left \{
V^{-1}(\Lambda_ P, k_1)\right \}\times\nonumber\\
&\times& D^{(S=1)}_{\nu_{2p}\nu_{2k}} \left\{ V^{-1} (\Lambda_{p_2},
k_2)\right \} D^{(S=1)}_{\nu_{2k}\nu_2}\left\{ V^{-1} (\Lambda_ P, k_2)\right
\},
\end{eqnarray}

\begin{equation}
V^{\nu_{1p}\nu_{2p}}_{\sigma_{1p}\sigma_{2p}} (\vec k(-) \vec p, \vec p) =
\xi_{\sigma_{1p}} \xi_{\sigma_{2p}} \hat V^{(2)} (\vec k(-) \vec p,
\vec p) \xi_{\nu_{1p}} \xi_{\nu_{2p}},
\end{equation}

Let us use the equations for the 4- current of spinor particle defined by the
formula (\ref{eq:current}) and Eq. (\ref{eq:cur}) for the 4- current of vector
particle in the Joos-Weinberg's formalism
\begin{equation}\label{eq:cur}
j^{\mu}_{\sigma_{2p}\nu_{2p}}(\vec p, \vec k)= \xi^{*}_{\sigma_{2p}}
\left [ (p_2+k_2)^\mu+\frac{1}{m_2}W^\mu(\vec p_2)(\vec S_2\vec
\Delta_2)-\frac{1}{m_2}(\vec S_2\vec\Delta_2)W^{\mu}(\vec p_2)
\right ]\xi_{\nu_{2p}}.
\end{equation}
Following to  the rules of construction of the quasipotential over the on-shell
scattering
amplitude~\cite{Logunov,Kadysh} we  obtain
\begin{eqnarray}
\lefteqn{<\vec p_1, \vec p_2; sigma_{1p}, \sigma_{2p}\mid V^{(2)}\mid \vec k_1,
\vec k_2;
\nu_{1p}, \nu_{2p}>=}\nonumber\\
&=&<\vec p_1, \vec p_2; \sigma_{1p}, \sigma_{2p}\mid T^{(2)} \mid \vec k_1,
\vec k_2;
\nu_{1p}, \nu_{2p}>= -g_V\frac{j^\mu_{\sigma_{1p}\nu_{1p}}(\vec p_1, \vec
k_1)g_{\mu\nu}j^{\nu}_{\sigma_{2p}\nu_{2p}}(\vec p_2,\vec k_2)}{(p_1-k_1)^2}
\end{eqnarray}
with , as earlier, $\vec p=\vec p_1=-\vec p_2$ and $\vec k=\vec k_1=-\vec k_2$
are covariant generalizations of momenta.

As a result one can write the quasipotential operator as follows
\begin{eqnarray}\label{eq:qua2}
\lefteqn{\hat V^{(2)}(\vec p, \vec k)= -
g_V^2\sqrt{\frac{\Delta_{1}^{0}+m_1}{2m_1}}
\frac{p_{1}^{0}(p_{1}^{0}+p_{2}^{0}+k_{1}^{0}+
k_{2}^{0})-2m^2_1}{m_1(\Delta_{1}^{0}-m_1)}-}
\nonumber\\
&-& g_V^2\frac{(p_{1}^{0}+p_{2}^{0}+k_{1}^{0}+k_{2}^{0})(\vec p
\vec\Delta_1)}{m_1(\Delta_{1}^{0}-m_1)
\sqrt{2m_1(\Delta_{1}^{0}+m_1)}}-g_V^2
\frac{i\vec \sigma_1 [\vec p\vec \Delta_1]
(p_{1}^{0}+p_{2}^{0}+k_{1}^{0}+k_{2}^{0})}{m_1(\Delta_{1}^{0}-m_1)
\sqrt{2m_1(\Delta_{1}^{0}+m_1)}}+\nonumber\\
&+&g_V^2\sqrt{\frac{\Delta_{1}^{0}+m_1}{2m_1}}\frac{i\vec S_2 [\vec p\vec
\Delta_2] (p_{1}^{0}+p_{2}^{0})}{m_1 m_2(\Delta_{1}^{0}-m_1)}+\nonumber\\
&+&g_V^2 \frac{1}{\sqrt{2m_1(\Delta_{1}^{0}+m_1)}}\frac{i\vec S_2 [\vec
p\vec\Delta_2](\vec p\vec \Delta_1)(p_{1}^{0}+p_{2}^{0}+m_1+m_2)^2}{2m_1
m_2(\Delta_{1}^{0}-m_1)(p_{1}^{0}+m_1)(p_{2}^{0}+m_2)}+\nonumber\\
&+&
g_V^2\sqrt{\frac{m_1}{2(\Delta_{1}^{0}+m_1)}}
\frac{(\vec\sigma_1\vec\Delta_2)(\vec S_2\vec\Delta_1)-(\vec\sigma_1\vec
S_2)(\vec\Delta_1\vec\Delta_2)
+i\vec S_2 [\vec\Delta_1\vec\Delta_2]}{m_1(\Delta_{1}^{0}-m_1)}-\nonumber\\
&-&g_V^2\frac{1}{\sqrt{2m_1(\Delta_{1}^{0}
+m_1)}}\frac{\vec \sigma_1 [\vec p\vec \Delta_1]\vec S_2 [\vec
p\vec\Delta_2](p_{1}^{0}+p_{2}^{0}+m_1+m_2)^2}{m_1
m_2(\Delta_{1}^{0}-m_1)
(p_{1}^{0}+m_1)(p_{2}^{0}+m_2)}.
\end{eqnarray}

Here,
\begin{eqnarray}
\vec\Delta_1&=&\vec k-\frac{\vec p}{m_1}\left (k_{1}^{0}-\frac{\vec k\vec
p}{p_{1}^{0}+m_1}\right ),\qquad \Delta_{1}^{0}=\sqrt{\vec
\Delta^2_1+m_1^2},\nonumber\\
\vec\Delta_2&=&\vec k-\frac{\vec p}{m_2}\left (k_{2}^{0}-\frac{\vec k\vec
p}{p_{2}^{0}+m_2}\right ),\qquad \Delta_{2}^{0}=\sqrt{\vec \Delta^2_2+m_2^2}.
\end{eqnarray}
and $p_{1}^{0}=\sqrt{\vec p\,^2+m_1^2}, k_{1}^{0}=\sqrt{\vec k\,^2+m_1^2},
p_{2}^{0}=\sqrt{\vec p\,^2+m_2^2}, k_{1}^{0}=\sqrt{\vec k\,^2+m_2^2}$.

In the quasirelativistic approximation (account of terms up to the order
$1/c^2$)  Eq. (\ref{eq:qua2}) yields
\begin{eqnarray}
\lefteqn{V^{(2)}_{nonrel.}(\vec k,\vec
p)=-g^2_V\frac{4m_1m_2}{\vec\Delta^2_{{\cal E}}}+g^2_V \frac{1}{c^2}\left
(1+\frac{m_2}{2m_1}\right )
-g^2_V \frac{1}{c^2}\left (2+\frac{m_1}{m_2}
+\frac{m_2}{m_1}\right )\frac{\vec p\,^2+\vec k\,^2}{\vec\Delta^2_{{\cal
E}}}-}\nonumber\\
&-&g_V^2 \frac{1}{c^2}\frac{m_2}{m_1}\frac{(\vec k\,^2-\vec p\,^2)^2}{\vec
\Delta_{\cal E}^4}-g_V^2 \frac{1}{c^2}\left (1+\frac{m_2}{m_1}\right
)\frac{2i\vec\sigma_1 [\vec p\vec\Delta^2_{{\cal E}}]}
{\vec\Delta^2_{\cal E}}-g_V^2 \frac{2}{c^2}\left (1+\frac{m_1}{m_2}\right
)\frac{i\vec S_2 [\vec p\vec\Delta_{\cal E}]}{\vec\Delta_{\cal
E}^2}-\nonumber\\
&-&g_V^2 \frac{1}{c^2}\frac{(\vec\sigma_1\vec\Delta_{\cal E})(\vec
S_2\vec\Delta_{\cal E})-(\vec\sigma_1\vec S_2)\vec\Delta_{\cal
E}^2}{\vec\Delta_{\cal E}^2},
\end{eqnarray}
where again $\vec\Delta_{\cal E}=\vec k-\vec p$ is the momentum transfer in the
Euclidian space.  As opposed to the Eq. (\ref{eq:nonrel}) we have two
additional terms corresponding to
the tensor forces and the spin-orbit interaction of the second particle.

One can see that  this case is more complicated comparing to the case of the
Section II and it does not admit the analytical solution.
Therefore, we intend to solve the equation with the quasipotential
(\ref{eq:qua2}) in the following publications numerically.
A good accuracy in numerical solution of   such a type of problems is provided
by the spline method~\cite{Sidorov} or  by the method for solving the spectral
problems, developed in Ref.~\cite{Zhidk}, and which is founded on the
Galerkin's procedure of discretization of integral operators. They  have been
used
for the description of two spinor system in~\cite{Arbuzov}.

\section{\bf Conclusions}

\hspace*{8mm}In  the  presented  paper  we  have  applied the covariant
three-dimensional quasipotential approach to the description of quark-diquark
bound states, which can be interpreted as nucleons and their resonances. We
have derived   the   partial
relativistic  equal-time  equation  for
($\pi\mu$)- atom  and  other bound  systems (e.g. proton)   composed   from
the
particles  with  spin  1/2 (quark)  and  spin  0 (diquark).
The spin structure of the quasipotential for the system of fermion and $S=1$
boson has been also under consideration.

The   presence   of  huge  terms   in    these
equations  induces us to  employ the  numerical   methods for their solution.\\

{\tt ACKNOWLEDGEMENTS.} We would like to express our sincere gratitude to V. G.
Kadyshevsky, R. N. Faustov, M. Moreno, N. B. Skachkov, Yu. N. Tyukhtyaev and C.
Villareal
 for interest in the work and most helpful discussions.
We greatly appreciate the technical assistance of A. S. Rodin.

One of us (V. D.) is most grateful to Prof. A. M. Cetto, Head of  the
Departamento de F\'{\i}sica
Te\'{o}rica at the IFUNAM,  for the creation of  excellent conditions
for work.

Finally, it  should be mentioned that this work has been financially supported
by the CONACYT (Mexico) under contract No. 920193
and Scientific and Technological Center for Control and Use of Physical Fields
and Radiations (Saratov) . \\
\\

{\large {\tt Appendix}}\\

First of all, let us mention that the integral  (\ref{eq:int})
\begin{equation}\label{eq:int26}
I^{(l)}_1=\int^1_{-1}\frac{P_l(z)dz}{(\gamma^{-}-z)\sqrt{\gamma^{+}-z}}
\end{equation}
can be reduced by means of simple algebraic transformations to
\begin{equation}\label{eq:int18}
I^{(l)}_1={1\over \gamma^{-}-\gamma^{+}}I^{(l)}_2+{2\over
\sqrt{\gamma^{+}-\gamma^{-}}}Q_l(\gamma^-) +{1\over \gamma^+
-\gamma^-}\int^{1}_{-1}\frac{P_l(z) dz }{\sqrt{\gamma^+ -z}+\sqrt{\gamma^+
-\gamma^-}},
\end{equation}
where $Q_l(x)$ is  the Legendre function of the second kind.
However, the calculation of the integral in (\ref{eq:int18}) is as
complicated as the previous one (\ref{eq:int26}).

Next,  we can use the  multiple Mellin
transform  and  the  tables  of  formulas  from  Ref.~\cite{Marich,Prud} in
order to  calculate
the  integral  (\ref{eq:int}).
The multiple Mellin transform
has the form:
\begin{equation}
K^{*}(s_1,\ldots, s_n)=\int^{\infty}_0\cdots \int^{\infty}_0K(x_1,\ldots, x_n)
x_1^{s_1-1}\ldots x_n^{s_n-1} dx_1 \ldots dx_n.
\end{equation}
If the function $K(c_1,..., c_n)$  can  be represented  in  such  a  form:
\begin{equation}
K(c_1,\ldots,c_n)=\int^{\infty}_0 K_1(x)K_2(\frac{c_1}{x})\ldots K_{n+1}
(\frac{c_n}{x})\frac{dx}{x}
\end{equation}
then  the  transform   $K(s_1,\ldots, s_n)$ is  calculated  by  the
formula
\begin{equation}
K^{*}(s_1,\ldots ,s_n) = K^*_1(s_1+\ldots + s_n)K^*_2(s_1)\ldots
K^*_{n+1}(s_n).
\end{equation}

After the  substitution $ z  =  2/x -1 $, the needed integral  is rewritten
as follows:
\begin{equation}\label{eq:int27}
I^{(l)}_1=\frac{(-1)^l \sqrt{\gamma^+ -1}}{\gamma^-
-1}\int^{\infty}_{0}\frac{dx}{x} P_l(\frac{2}{x}-1)H(x-1)\cdot\frac{c/x}
{\sqrt{1+c/x}(1+\frac{\gamma^+ -1}{\gamma^- -1}\cdot c/x)},
\end{equation}
where  $c = 2/(\gamma^{+}-1)$, $H(y) = \cases{1, y\ge 0 ;
\cr 0, y<0 \cr}$ is the Heviside function.

Taking the transforms from the Tables~\cite{Prud,Bateman}
we can use the reverse Mellin transformation to find out the value of the
integral (\ref{eq:int27})
\begin{equation}\label{eq:rmel}
K(c_1,\ldots, c_n)=\frac{1}{(2\pi
i)^n}\int^{\gamma_1+i\infty}_{\gamma_1-i\infty}
\cdots\int^{\gamma_n+i\infty}_{\gamma_n-i\infty}K^{*}(s_1,\ldots,s_n)
c_1^{-s_1}\ldots c_n^{-s_n} ds_1\ldots ds_n,
\end{equation}
where $\gamma_k = \Re e\, s_k , k=1,\ldots n$.

Thus,
\begin{eqnarray}
\lefteqn{I^{(l)}_1=(-1)^{l}\frac{\sqrt{\gamma^+ -1}}{\gamma^-
-1}\sum^{\infty}_{k=0}\frac{1}{\Gamma({3\over 2}+k)}\left (\frac{\gamma^-
-\gamma^+}{\gamma^- -1}\right )^k\times}\nonumber\\
&\times&G^{1,3}_{3,3}\left ( \frac{2}{\gamma^+ -1}\matrix{
\mid&1, & 1, & {1\over 2} \cr
\mid&1+k, & -l, & 1+l \cr
}\right ),
\end{eqnarray}
where $G^{A,B}_{B+C, A+D}$ is the Meijer $G$- function.

The another way is also possible: to consider every term  in the integral
(\ref{eq:int26}) separately and to employ triple\footnote{The number of
multiple terms in the integral (\ref{eq:int26})  is three.} Mellin transforms.
Using  this  technique we obtain~\cite{Marich}
\begin{equation}
K_1(x)=P_l(\frac{2}{x}-1)H(x-1)\,\,\Rightarrow\,\,K_1^{*}(s)=
\Gamma\Bigl[{-s,-s\atop l+1-s,-l-s}\Bigr],\,
\end{equation}
\begin{equation}
K_2(\frac{c_1}{x})=\frac{1}{\frac{x}{c_1}-1}\,\,
\Rightarrow\,\,K_2^{*}(s_1)=-\pi
\Gamma\Bigl[{-s_1,1+s_1\atop\frac{1}{2}-s_1,\frac{1}{2}
+s_1}\Bigr],
\end{equation}
\begin{equation}
K_3(\frac{c_2}{x})=\frac{1}{\sqrt{1-\frac{c_1}{x}}}\,\,\Rightarrow\,\,
K_3^{*}(s_2)=\frac{\pi}{\Gamma(\frac{1}{2})\cos\frac{\pi}{4}}
\Gamma\Bigl[{s_2,\frac{1}{2}-s_2\atop\frac{1}{4}+s_2,\frac{3}{4}-s_1}\Bigr],
\end{equation}
where $c_1={2\over \gamma^- +1}$, $c_2={2\over \gamma^+ +1}$ and $s=s_1+s_2$,
$\Gamma(s)$ is the  Eiler's $\Gamma$-  function  and
$\Gamma\Bigl[{a_1\ldots a_k \atop b_1\ldots b_m}\Bigr]$ denotes

$$\Gamma\Bigl[{a_1\ldots a_k \atop b_1\ldots b_m}\Bigr]=\frac{\Gamma(a_1)
\ldots \Gamma(a_k)}{\Gamma(b_1)\ldots \Gamma(b_m)}.$$

Then,
\begin{equation}
K^{*}(s_1,s_2)=-\frac{\sqrt{2}\pi^2}{\Gamma(\frac{1}{2})}
\Gamma\Bigl[{-s_1-s_2,-s_1-s_2,-s_1,1+s_1,s_2,\frac{1}{2}-s_2\atop
l+1-s_1-s_2,-l-s_1-s_2,\frac{1}{2}-s_1,\frac{1}{2}+s_1,\frac{1}{4}+s_2,
\frac{3}{4}-s_2}\Bigr].
\end{equation}
Similarly to the preceding calculation,  employing the reverse Mellin
transformation (\ref{eq:rmel})
\begin{equation}
K(c_1,c_2)=\frac{1}{(2\pi i)^2}\int^{\gamma_1+i\infty}_{\gamma_1-i\infty}
\int^{\gamma_2+i\infty}_{\gamma_2-i\infty}K^{*}(s_1,s_2)
c_1^{-s_1}c_2^{-s_2}ds_1 ds_2
\end{equation}
to our integral we come to
\begin{eqnarray}
\lefteqn{K(c_1,c_2)=-\frac{\sqrt{2}\pi^2}{\Gamma(\frac{1}{2})}
\sum_{k=0}^{\infty}
\sum_{n=0}^{\infty}
\frac{(-1)^{k+n}}{k!n!}\times}\nonumber\\
&\times &\Gamma\Bigl[{k+n+1,k+n+1,k+1,
\frac{1}{2}+n
\atop l+k+n+2,-l+k+n+1,\frac{3}{2}+k,-\frac{1}{2}-k,
\frac{1}{4}-n,
\frac{3}{4}+n}\Bigr]c_1^{k+1}c_2^{n},
\end{eqnarray}
which can be slightly simplified after  use  of well--known expressions for
$\Gamma$- function,\\
$\Gamma(p) \Gamma(1~-~p)~=~\frac{\pi}{\sin p\pi}$, and\,\,
$\Gamma(k+1)=k!\quad (k=0,1\ldots)$

Finally,  the value of the  integral (\ref{eq:int26}) can be represented  in
the form of  the  complicated  double sum of $\Gamma$- functions:
\begin{equation}
I^{(l)}_1=\frac{1}{\sqrt{\pi(\gamma^{+}+1)}}
\sum_{k=0}^{\infty}\sum_{n=0}^{\infty}
\Gamma\Bigl[{k+n+1, k+n+1, \frac{1}{2}+n
\atop l+k+n+2, -l+k+n+1, n+1}
\Bigr]c_1^{k+1}c_2^{n}.
\end{equation}

It is not clear, what representation of the  integral under consideration is
more convenient; all of them have enormous form and are rather inconvenient
for some applications.
In our opinion, the further simplifications appear to be impossible.
Therefore,  the use of computer seems  to be necessary.
\newpage

\vspace*{5mm}
{\large{\tt References}}
\vspace*{-5mm}

\end{document}